\documentclass[preprint]{aastex}

\usepackage{graphics}
\newcommand{\heI}{He~\textsc{i} }
\newcommand{\hI}{H~\textsc{i} }
\newcommand{\av}{A_{\rm V}}
\newcommand{\rv}{R_{\rm V}}

\shorttitle{IR-Excess Stellar Objects in G54.1+0.3}
\shortauthors{Kim, Koo, and Moon}

\begin{document}
\title{Near-Infrared Spectroscopy of Infrared-Excess Stellar Objects 
in the Young Supernova Remnant G54.1+0.3}

\author{Hyun-Jeong Kim\altaffilmark{1} and Bon-Chul Koo\altaffilmark{2}}
\affil{Department of Physics and Astronomy, 
Seoul National University,\\ 
Seoul 151-742, Republic of Korea}

\and
 
\author{Dae-Sik Moon\altaffilmark{3}}
\affil{Department of Astronomy and Astrophysics, 
University of Toronto,\\
Toronto, ON M5S 3H4, Canada \\
Visiting Brain Pool Scholar, Korea Astronomy and Space Science Institute,\\
Daejeon 305-348, Republic of Korea}
\altaffiltext{1}{hjkim@astro.snu.ac.kr}
\altaffiltext{2}{koo@astro.snu.ac.kr}
\altaffiltext{3}{moon@astro.utoronto.ca}

\begin{abstract}
We present the results of broadband near-infrared spectroscopic 
observations of the recently discovered mysterious stellar objects 
in the young supernova remnant G54.1+0.3. 
These objects, which show significant mid-infrared excess emission,
are embedded in a diffuse loop structure of $\sim$1$\arcmin$ in radius.
Their near-infrared spectra reveal characteristics of late O- or early B-type stars 
with numerous H and \heI absorption lines, 
and we classify their spectral types to be between O9 and B2 
based on an empirical relation derived here 
between the equivalent widths of the H lines and stellar photospheric temperatures.
The spectral types, combined with the results of spectral energy distribution fits, 
constrain the distance to the objects to be 6.0 $\pm$ 0.4 kpc.
The photometric spectral types of the objects are consistent with
those from the spectroscopic analyses, 
and the extinction distributions indicate 
a local enhancement of matter in the western part of the loop. 
If these objects originate via triggered formation by the progenitor star of G54.1+0.3,
then their formations likely began during the later evolutionary stages of 
the progenitor, although a rather earlier formation may still be possible.
If the objects and the progenitor belong to the same cluster of stars,
then our results constrain the progenitor mass of G54.1+0.3 to be 
between 18 and $\sim$35\,$M_{\sun}$ and 
suggest that G54.1+0.3 was either a Type IIP supernova or,
with a relatively lower possibility, Type Ib/c from a binary system.
\end{abstract}


\section{Introduction} 

Young core-collapse supernova remnants (SNRs), 
especially those with a connate pulsar,
are usually rich in features useful 
for studying a diverse range of astrophysical problems.
The very existence of a pulsar verifies 
the core-collapse nature of a progenitor star and 
limits the age of, and often distances to, an SNR,
providing basic key information for better understanding of 
other relevant phenomena 
such as the evolution and dynamics of pulsar wind nebulae
(PWNe) and supernova (SN) ejecta, distribution of circumstellar
material from the mass loss of a progenitor, and shock interactions
with the interstellar medium (ISM) \citep[e.g.,][]{weisskopf00,
gaensler02, moon04, koo07, lee09, moon09}.
As we describe here,
G54.1+0.3 is one such SNR that provides a unique opportunity for the
in-depth study of either star formation triggered by an SN progenitor 
or the evolution of SN dust ejecta.

G54.1+0.3 is a young, core-collapse SNR with a central PWN that is
prominent in synchrotron radio and X-ray emissions. As this SNR
closely resembles the Crab nebula, it is often referred to as a
cousin of the Crab nebula \citep{velusamy88,lu02}. Its central
pulsar (PSR J1930+1852), which is surrounded by a faint shell-like
radio emission of $\sim$8$\arcmin$ in diameter \citep{lang10}, has a
136-ms rotational period and a characteristic age of 2,900 yrs
\citep{camilo02,lu02}.
There is faint X-ray emission distributed
along the inner boundary of the radio shell \citep{bocchino10}, and
together the radio and X-ray morphologies strongly suggest that the
emissions are from an SNR shell propagating into an ambient medium of
an interstellar and/or circumstellar origin.
Therefore, G54.1+0.3
belongs to a category of composite SNRs showing both a central PWN
(or plerion) and a surrounding SNR shell. Notably, the recent
detection of strong $\gamma$-ray emission from the SNR renders its
PWN to be one with the highest $\gamma$-ray to X-ray luminosity
ratios among all known PWNe driven by young rotation-powered pulsars
\citep{acciari10}.
\hI and CO observations give the kinematic distance to G54.1+0.3 
as 5--10 and 6--8 kpc, respectively \citep{koo08,leahy08}, 
while the dispersion measure of its central pulsar places 
the SNR at a distance of $\sim$9 kpc \citep[][see \S~3.3]{cordes02}.

Interest in G54.1+0.3 has recently been sparked by the discovery of
its diffuse infrared (IR) loop with a $\sim$1$\arcmin$ radius
\citep{koo08}. The IR loop, which was identified by observations
using the {\em AKARI} satellite, consists of at least 11 embedded
stellar objects with significant excess emission in the mid-infrared
(MIR) wavebands; here, we refer to these as ``IR-excess stellar
objects.''
Their IR colors and spectral energy distributions (SEDs)
are consistent with those of massive ($\gtrsim$10\,$M_{\sun}$)
pre-main sequence stars with destroyed inner disks, which
subsequently leads to the interpretation that their formation was
triggered by a progenitor star of the SN in G54.1+0.3 \citep{koo08}.
The {\it Spitzer} spectroscopy of the IR loop, however, implies that
it is more likely made of freshly formed dust produced
by the ejecta of the SN explosion and that the IR-excess stellar
objects are in fact early-type main sequence (MS) stars of a stellar
cluster to which the progenitor star originally belonged \citep{temim10}.
Thus, it is imperative to carry out near-infrared (NIR) spectroscopic 
observations of the IR-excess stellar objects to obtain more direct information 
for classifying the spectral types of the IR-excess stellar objects 
and, eventually, the origin of the IR loop.

In this paper, we present the results of a NIR spectroscopy study of
the IR-excess stellar objects and discuss the implications of the
results in the context of the two possible scenarios for their
origin. The paper is organized as follows: in \S~2, we describe our
observations and data reduction, followed by the spectral analyses
and classifications as well as the distance determination in \S~3.
In \S~4, we compare the results of the spectroscopic spectral
classification with those from SED fit analyses and investigate the
extinction distribution around the IR-excess stellar objects. We
then discuss the origin of the IR-excess stellar objects in \S~5 and
give our summary and conclusions in \S~6.

\section{Observations and Data Reduction}

\subsection{Observations}

NIR spectroscopic observations were performed for seven of the 11
IR-excess stellar objects discovered by \citet{koo08} in the SNR
G54.1+0.3 with the TripleSpec spectrograph on the 5-m Palomar Hale
telescope on 2008 August 9. TripleSpec is a slit-based NIR
cross-dispersion echelle spectrograph covering the entire NIR
atmospheric window simultaneously with a spectral resolving power
$R$ of 2500--3000 using six (from 3 to 8) echelle orders
mapped onto two adjacent quadrants of a Hawaii II HgCdTe 2K $\times$
2K detector array from Teledyne Inc. \citep{wilson04,herter}. The
slit width and length of TripleSpec are 1\arcsec\ and 30\arcsec,
respectively, and the broad spectral coverage of the spectrograph
makes it ideal for conducting observations of new objects with
little known information such as the IR-excess stellar objects in
G54.1+0.3.

\begin{figure}
\center
\includegraphics[width=0.6\textwidth]{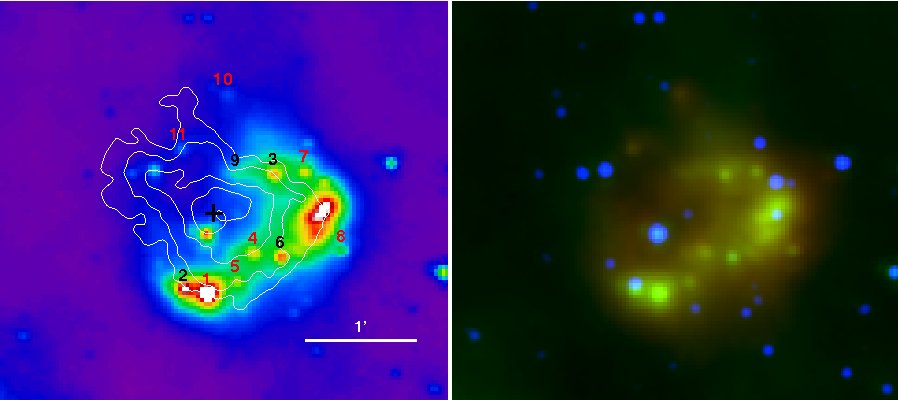}
\caption{{\it Left:} {\em AKARI} 15~$\micron$ image of the SNR G54.1+0.3 
and the 11 IR-excess stellar objects within the IR loop. 
The cross indicates the position of the pulsar at 
($\alpha, \delta) = (19^{\rm h}30^{\rm m}30^{\rm s}.13, 
+18\arcdeg52\arcmin14\arcsec.1$; J2000.0). 
The VLA 4.85 GHz radio contours are also shown as an overlay.
The contour levels are equally spaced 
every 1.28 K from 0.22 K in the brightness temperature and 
increase toward the center of the SNR. 
The numbers of the IR-excess stellar objects are 
in order of the {\it Spitzer} MIPS 24~$\micron$ 
brightness determined by PSF photometry \citep{koo08}. 
Among them, the NIR spectra of seven objects marked with red 
were obtained in this study. 
The scale bar corresponds to 1\arcmin. North is up and east is to the left. 
{\it Right:} Three-color image generated from the {\it Spitzer} 
IRAC 5.8~$\micron$ (B), the {\em AKARI} 15~$\micron$ (G), 
and the {\it Spitzer} MIPS 24~$\micron$ (R). \label{fig1}}
\end{figure}

Figure~\ref{fig1} shows the positions of the IR-excess stellar
objects in the {\em AKARI} 15~$\micron$ image numbered in decreasing
order of their {\it Spitzer} MIPS 24~$\micron$ brightness \citep{koo08}. 
Table~\ref{tbl1} lists the coordinates and magnitudes in the visible 
to MIR wavebands apart from those for Object 2, which has been detected 
only at $\lambda \gtrsim 3.6\, \micron$ \citep{usno,nomad,koo08}.
The photometric data used in \citet{koo08} are based on the {\it Spitzer} 
IRAC Galactic Legacy Infrared Mid-Plane Survey Extraordinaire (GLIMPSE)%
\footnote{http://www.astro.wisc.edu/glimpse/glimpsedata.html} catalog,
which somehow does not contain the 2MASS NIR magnitudes of Object 6; 
these are instead obtained from the 2MASS All-Sky Point Source Catalog%
\footnote{http://www.ipac.caltech.edu/2mass/releases/allsky}
\citep{2mass} and added in Table~\ref{tbl1}.

The seven objects (i.e., 1, 4, 5, 7, 8, 10, and 11) marked in red in
Figure~\ref{fig1} were observed with TripleSpec in this work.
However, the low signal-to-noise (S/N) ratio of Object 4, the
faintest one in the {\it K}$_{s}$-band, makes the reliable use of
its spectrum difficult, and consequently, it has been excluded from
further analysis. All the objects were observed with the typical AB
nodding pattern, separated by 10$\arcsec$ along the slit direction,
to effectively subtract the sky background emission. The total
exposure time of each object was 600 s. The standard star HD 171623
(A0V) was observed at similar airmasses to those of the objects for
standard star calibration.

\subsection{Data Reduction}

\begin{figure}
\center
\includegraphics[width=0.5\textwidth]{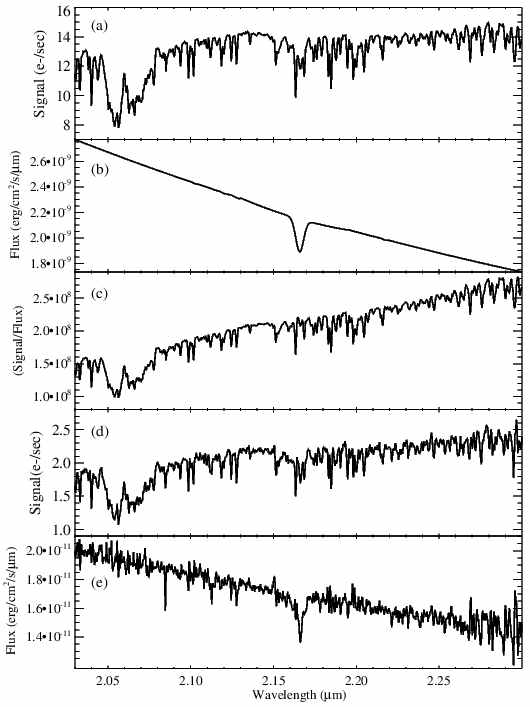}
\caption{Standard star calibration process.
{\bf (a)} The observed spectrum of the standard star HD 171623. 
{\bf (b)} The model spectrum of HD 171623. 
{\bf (c)} The telluric line spectrum including the TripleSpec system
responsivity obtained by dividing (a) by (b). 
{\bf (d)} The observed spectrum of Object 1. 
{\bf (e)} The final flux-calibrated spectrum of Object 1.
\label{fig2}}
\end{figure}

We first conduct flat fielding and bad pixel correction before
obtaining two-dimensional wavelength solutions of the dispersed
images using the OH sky emission lines. For each frame of a dithered
pair, we subtract the sky background of a given frame using the
other frame and then combine the pair of sky-subtracted spectra from
the two frames to produce the final spectrum. Although TripleSpec
covers wavelengths from 0.78 to 2.4~$\micron$ using six echelle
orders of 3--8, we only use orders 3--6 (i.e., $\sim$1.0--2.4~$\micron$) 
due to the low S/N ratios of the spectra obtained in orders 7 and 8. 
The S/N ratios of the final spectra range between 5 and 30. 
The extracted spectra show the features of the telluric absorption lines 
convolved with the TripleSpec system responsivity (Figure~\ref{fig2}~(a) and (d)).

For the standard source calibration, we are unable to find an appropriate 
standard spectrum of our calibration source HD 171623 (A0V) 
that can be used reliably to calibrate our TripleSpec spectra. 
Instead, we use the scaled model spectrum of Vega, 
which has the same spectral type and luminosity class as HD 171613,
generated by the ATLAS12 stellar models \citep{castelli} as the
standard spectrum of our calibration source.
The widths of the model H absorption lines of the Vega spectrum are 
larger than those of the observed lines of HD 171623; therefore, 
we replace those lines, 
including Br$\gamma$, Br11, Br12, Pa$\beta$, and Pa$\delta$, 
with a Voigt profile in a scaled model spectrum if the lines do not blend or overlap. 
In the case where the H and telluric lines are adjacent (e.g., Br$\gamma$), 
the telluric lines are modeled using a Gaussian profile and subtracted.

Figure~\ref{fig2} presents an example of this calibration procedure
for the order-3 (2.03--2.40~$\micron$) spectrum of Object 1. 
The observed spectrum of the standard source HD 171623 in Figure~\ref{fig2}~(a) 
exhibits many telluric absorption lines and broad CO$_{2}$ absorption bands 
near 2.06~$\micron$. This is divided by the model spectrum (Figure~\ref{fig2}~(b)) 
to produce the spectrum in Figure~\ref{fig2}~(c) featuring the telluric lines 
and system responsivity. The final calibrated spectrum (Figure~\ref{fig2}~(e))
is obtained by dividing the observed spectrum (Figure~\ref{fig2}~(d)) 
by the spectrum in Figure~\ref{fig2}~(c).

\section{Spectral Type Classification and Distance Determination}

\subsection{Near-Infrared Spectra of Infrared-Excess Stellar Objects}

\begin{figure}
\includegraphics[width=0.9\textwidth]{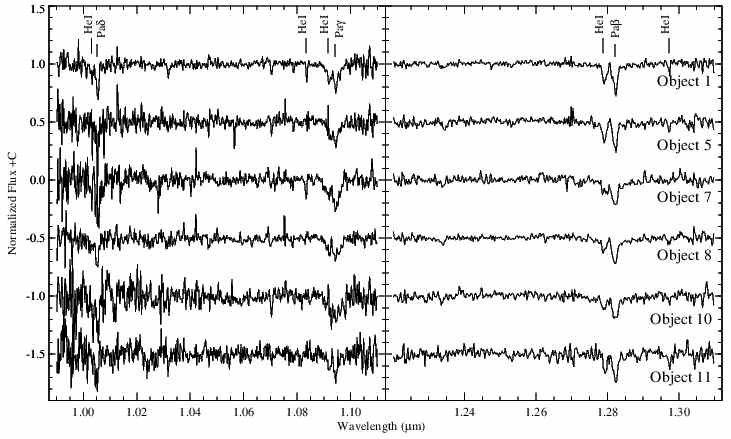}
\caption{Orders 6 and 5 of the spectra of all the IR-excess stellar
objects (Object 4 with a very low S/N ratio is excluded). 
The spectra are normalized by a 3rd- or 4th-order polynomial.
\label{fig3}}
\end{figure}

\begin{figure}
\includegraphics[width=0.9\textwidth]{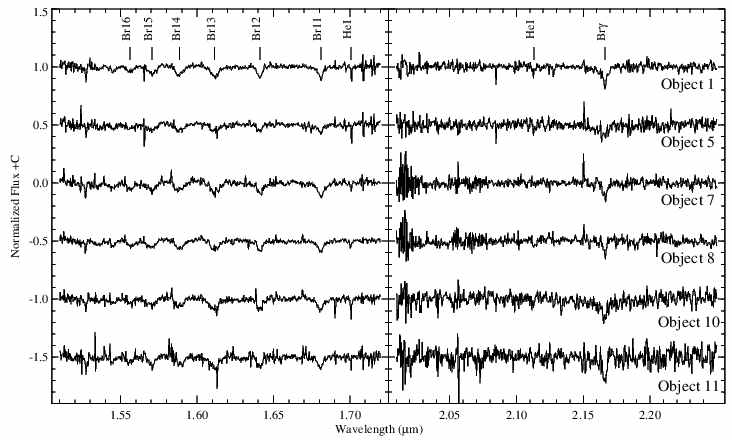}
\caption{Same as Fig.~\ref{fig3}, but for orders 4 and 3. \label{fig4}}
\end{figure}

Figures~\ref{fig3} and~\ref{fig4} present the obtained spectra of
the IR-excess stellar objects after normalization by a 3rd- or 4th-order polynomial 
in wavelength ranges of 0.99--1.11, 1.22--1.31, 1.51--1.72, and 2.01--2.25~$\micron$, 
corresponding to orders 6, 5, 4, and 3, respectively. 
Since higher orders have low throughputs for grating-based spectroscopy, 
the order-6 spectra have lower S/N ratios than the others. 
Also note that the order-3 spectra are heavily contaminated 
by numerous telluric lines. 
For all the spectra, photospheric absorption lines are dominant; 
in particular, the H Brackett series are pronounced, 
including six continuous transitions of the Br11--16 series. 
In increasing order of the wavelength, 
we identify the following absorption lines: 
\heI 1.003, Pa$\delta$ 1.005, \heI 1.083, \heI 1.092, and Pa$\gamma$
1.094~$\micron$ (order 6; Figure~\ref{fig3} {\it left});
\heI 1.279, Pa$\beta$ 1.282, and \heI 1.297~$\micron$
(order 5; Figure~\ref{fig3} {\it right}); 
Br16 1.556, Br15 1.571, Br14 1.589, Br13 1.611, Br12 1.641, 
Br11 1.681, and \heI 1.701~$\micron$ (order 4; Figure~\ref{fig4} {\it left}); 
and \heI 2.113 and Br$\gamma$ 2.166~$\micron$ 
(order 3; Figure~\ref{fig4} {\it right}).

We measure the equivalent widths ({\it EW}s) of the identified lines 
with a high S/N ratio by fitting a Gaussian profile, and these
results are presented in Table~\ref{tbl2}. 
Note that the {\it EW}s of the \heI 1.297~$\micron$ line of 
Object 11, Br11 of Object 7, and Br$\gamma$ of Object 5 are large 
because of contamination by nearby telluric features. 
In the obtained spectra shown in Figures~\ref{fig3} and~\ref{fig4}, 
we have not identified any obvious emission lines. 
Some emission-like features in the spectra 
(e.g., those around $\sim$1.27 and $\sim$1.57~$\micron$) are
residuals of strong telluric lines that have not been completely
removed during the standard star calibration.

\subsection{Spectral Type Classification}

The obtained spectra (Figures~\ref{fig3} and \ref{fig4}) of the
IR-excess stellar objects show absorption lines of
strong H transitions and relatively weak \heI transitions. 
This, together with the lack of He~\textsc{ii} lines, 
indicates that these are stars of a late O- and early B-type 
between O8 and B3 \citep{hanson96, hanson98, hanson05}. 
This is consistent with the previous interpretation based on 
the {\it JHK}$_s$ color-color diagram and SEDs of these objects \citep{koo08}.
Also, the absence of the \heI 2.059~$\micron$ and Br$\gamma$ emission lines 
suggests that they are MS stars rather than supergiants \citep{hanson96}.

To determine their spectral types more precisely, we first conduct 
a thorough literature search of reliable stellar spectroscopic libraries. 
However, as far as we are aware, only limited studies have been performed 
for NIR spectroscopic studies of OB stars so far,
and even those available studies are somewhat inadequate for 
spectral classification use due to their small sample sizes
\citep[e.g.,][]{wallace97, meyer98, wallace00} or paucity of the
identified lines useful for line ratio comparison
\citep[e.g.,][]{hanson96, hanson98, hanson05, bik05}.

\begin{figure}
\center
\includegraphics[width=0.55\textwidth]{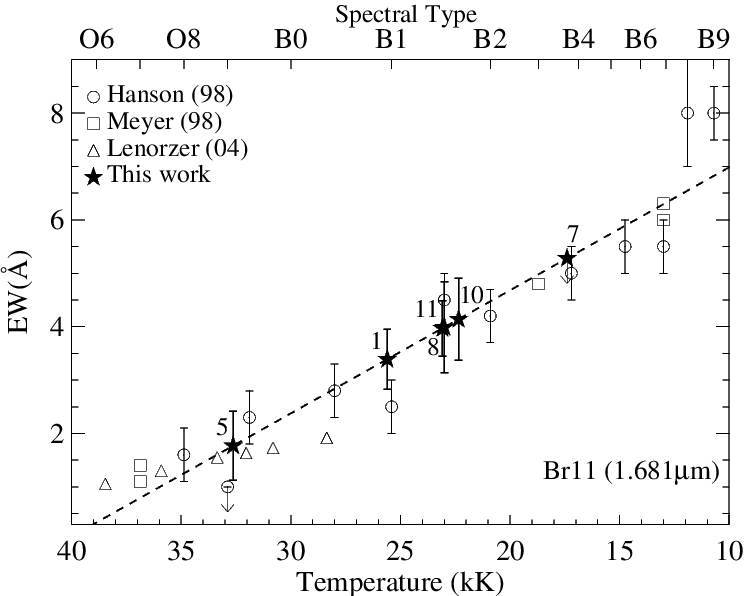}
\caption{The relation between the {\it EW}s of Br11 and 
the stellar temperatures.
{\it Open circles, boxes, and triangles} are from 
\cite{hanson98}, \cite{meyer98}, and \cite{lenorzer04}, respectively. 
The dashed line is from a linear fitting of the {\it EW}s 
to the temperatures from \cite{hanson98} ({\it open circles}).
{\it Filled stars} show the positions of the six IR-excess stellar objects 
in this study on this relation. \label{fig5}}
\end{figure}

\begin{figure}
\center
\includegraphics[width=0.55\textwidth]{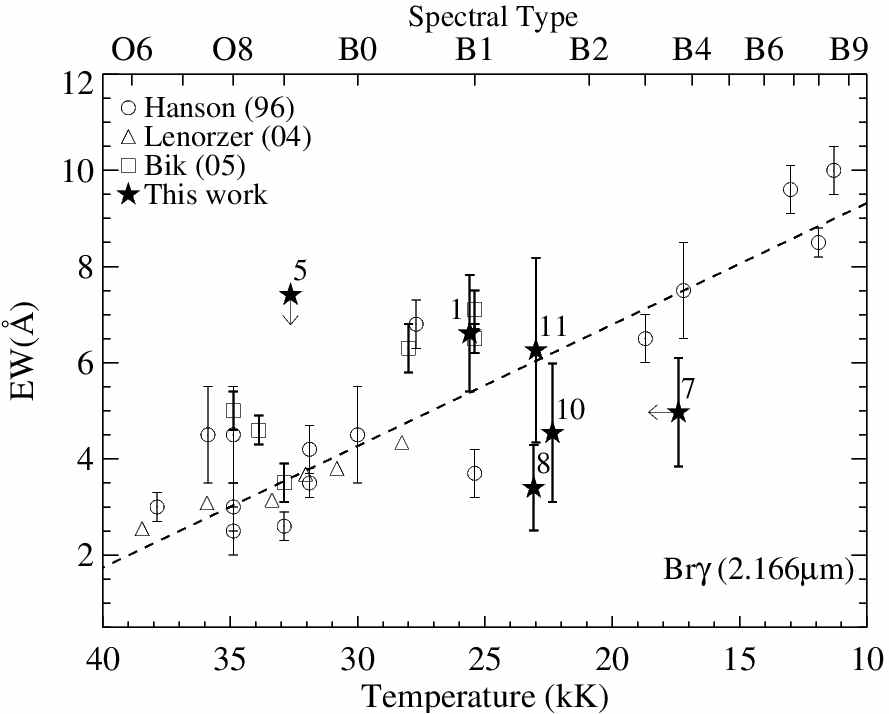}
\caption{Same as Fig.~\ref{fig5}, but for Br$\gamma$. 
{\it Open circles, triangles, and boxes} are from 
\cite{hanson96}, \cite{lenorzer04}, and \cite{bik05}, respectively. 
{\it Filled stars} are the IR-excess stellar objects of spectral types 
adopted from the results of Br11, i.e., determined by Eqn.~\ref{eqn1}. 
The dashed line is a linear fit of \cite{hanson96} 
({\it open circles}).\label{fig6}}
\end{figure}

The photospheric H lines of OB stars, on the other hand, 
are known to be very sensitive to temperature \citep{lenorzer04}, 
and, consequently, the {\it EW}s of the H lines are expected to be 
correlated with the stellar temperatures. 
Figure~\ref{fig5} confirms a correlation between the two parameters 
in the case of Br11, which exhibits a tight correlation 
(with a correlation coefficient of $-$0.94) of
\begin{equation}
EW_{Br11}[\AA] =
-0.23(\pm 0.02) \times (T\, {\rm [kK]})+9.29(\pm0.46) \label{eqn1}.
\end{equation}
Here, we use the results of \citet{martins05} and \citet{schmidt}
for the temperatures of OB stars and those of \citet{hanson98} for
the line widths. The results of \citet{meyer98}, which are based on
much smaller sample sizes, also agree well with this relation. Note
that among all the H lines in Table~\ref{tbl2} Br11 is an ideal
choice for this type of analysis because it is relatively well
isolated and thereby free of contamination from nearby sky/telluric
lines, although other photospheric H lines may be used for the same
purpose as long as reliable measurements of their {\it EW}s are
made. For instance, Figure~\ref{fig6} compares the temperatures of
OB stars and the {\it EW}s of their Br$\gamma$ lines available from
\citet{hanson96}, where the two parameters are still well correlated as
\begin{equation}
EW_{Br\gamma}[\AA] =
-0.25(\pm 0.01) \times (T\, {\rm [kK]})+11.84(\pm0.33) \label{eqn2}.
\end{equation}
The increased scatter in Figure~\ref{fig6} between the two
parameters compared with that in Figure~\ref{fig5} is mainly caused
by larger errors in the measured {\it EW}s of the Br$\gamma$ lines,
which are near telluric lines. The temperatures of the IR-excess
stellar objects obtained using Eqn.~\ref{eqn1} are presented in
Table~\ref{tbl3} along with their spectral types ranging between O9
and B2.

The uncertainties in the estimated temperatures of the IR-excess
stellar objects in Table~\ref{tbl3} (column 2) represent those from
the measurement errors of the {\it EW}s of the Br11 lines. The
scatter in the correlation between the {\it EW}s and the stellar
temperatures in Eqn.~\ref{eqn1} is an additional source of
uncertainty of $\sim$2,000~K. The combined uncertainties of these
two are roughly equivalent to one subclass in the determined
spectral types (column 3) of the IR-excess stellar objects.
Furthermore, there is an intrinsic difference of approximately one
spectral subclass around the late O and early B spectral types
between the two stellar models of \citet{martins05} and
\citet{schmidt}. We therefore believe that the spectral types of the
IR-excess stellar objects in Table~\ref{tbl3} have uncertainties
equal to a total of two subclasses.

It is notable that the measured {\it EW}s ($\sim$1~\AA) of the \heI\
lines (Table~\ref{tbl2}) are also consistent with the interpretation
that the spectral types of the IR-excess stellar objects lie between
late O- and early B-type stars \citep[see][for details]{hanson96,
hanson98}.

\subsection{Distance Determination}

We estimate the photometric distance to the six IR-excess stellar
objects based on the determined temperatures by fitting their SEDs
with the ATLAS9 Kurucz stellar models \citep{castelli03}. For this,
we use wavelengths shorter than 5~$\micron$ in the SED fits, to
concentrate on stellar emission, and the interstellar reddening law
of $\rv$ = 3.1 \citep{draine03}.
The best-fit distances and extinctions lie within 3.3--8.1 kpc 
and $\av$ = 6.9--7.8 mag, respectively, with mean values 
of 6.0 $\pm$ 0.4 kpc and 7.3 $\pm$ 0.1 mag (Table~\ref{tbl3}).

\begin{figure}
\center
\includegraphics[width=0.45\textwidth]{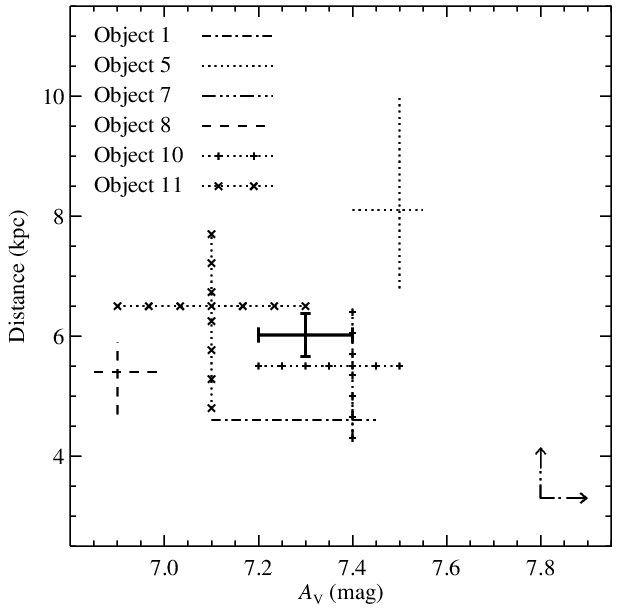}
\caption{Distances and extinctions to the individual IR-excess stellar 
objects derived from the SED fitting with the upper and lower temperature 
limits determined by Eqn.~\ref{eqn1}.
The mean distance $6.0 \pm 0.4$ kpc and mean extinction 
$\av=$ 7.3 $\pm$ 0.1 mag are shown as solid lines.
Object 7 only gives lower limits (see \S~3.3). \label{fig7}}
\end{figure}

Figure~\ref{fig7} shows the distribution of the obtained distances
and extinctions. The two parameters of Object 7 only have lower
limits because the lack of a $B$-band magnitude makes it difficult
to determine the upper limits---note that the $B$-band magnitude is
critical in determining the temperature of OB-type stars in the SED
fits. The large extinction ($\av \gtrsim$ 7.8 mag) together with the
absence of the $B$-band magnitude indicates that Object 7 has a
locally increased extinction. We discuss the extinction of the
IR-excess stellar objects in \S~4.

The distance to G54.1+0.3 was previously determined in several 
studies to be 5--10 kpc based on 21~cm H~\textsc{i} absorption lines 
\citep{koo08, leahy08} and $9^{+1.0}_{-1.5}$ kpc from a dispersion
measure analysis of the central pulsar J1930+1852 \citep{camilo02,cordes02}.
Two additional distances of 8 kpc and 6.2$^{+1.0}_{-0.6}$ kpc 
have been proposed from $^{13}$CO $J=$ 1--0 observations of 
nearby molecular clouds \citep{koo08, leahy08},
although it is uncertain whether the clouds are indeed associated
with the SNR \citep{lee12}. The 6.0 $\pm$ 0.4 kpc distance to the
IR-excess stellar objects derived from the SED fits in this study is
somewhat smaller but still consistent with previously reported values. 
Since our result is independent of the Galactic rotation model, 
we adopt 6 kpc as the distance to both the IR-excess stellar
objects and G54.1+0.3 in the following discussion.

\section{Photometric Spectral Classification and Extinction Distribution}

As we have shown, the spectral features of the six observed 
IR-excess stellar objects in the IR loop of the SNR G54.1+0.3 indicate that 
they are late O- or early B-type stars, which is consistent with 
the previous interpretation based on their {\it JHK}$_s$ colors \citep{koo08}. 
Given the importance of the spectral classification of 
the IR-excess stellar objects (see \S~1 and 5) in understanding their nature 
in association with G54.1+0.3, it is
worthwhile obtaining spectral classifications of the remaining five
objects, i.e., Objects 2, 3, 4, 6, and 9. For this, we again fit
SEDs in the $\sim$0.4-to-5-$\micron$ range of all the IR-excess
stellar objects using the ATLAS9 Kurucz stellar models \citep{castelli03}. 
We fix the distance to 6 kpc (as determined in \S~3.3) 
and derive the best-fit temperatures and extinctions, which
allows us to obtain photometric spectral classifications of these
five objects lacking NIR spectroscopic information. For the other
six objects classified using the NIR spectra (\S~3.2), the SED fits
provide us with confirmation that the derived spectral types are
correct. Object 2, however, is excluded from the SED fits because it
is detected only at longer ($\gtrsim$ 3 $\micron$) wavebands. (Our
most recent observations reveal that this object is composed of two
sources [H.-J. Kim et al. in preparation].)

\begin{figure}
\center
\includegraphics[width=0.8\textwidth]{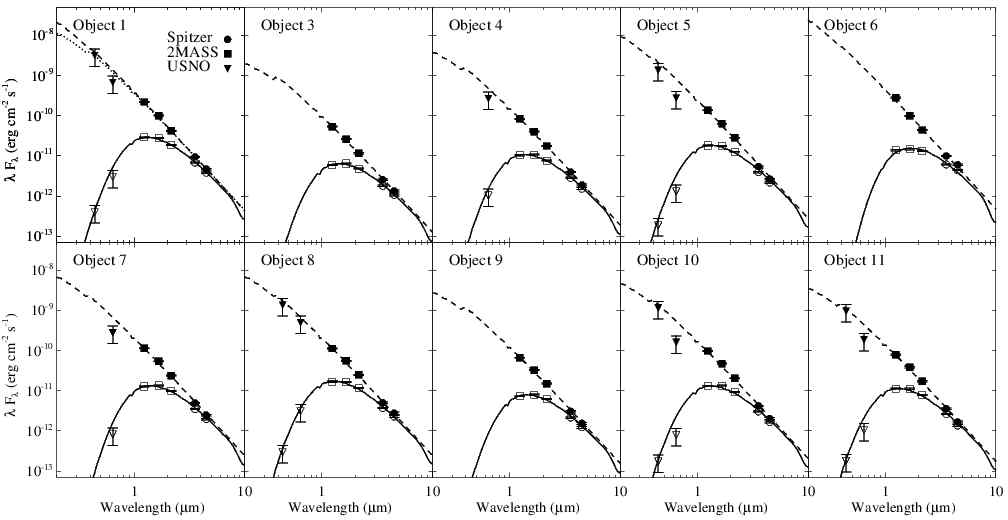}
\caption{Observed and best-fit SEDs of the IR-excess stellar objects. 
The open symbols and solid lines represent the observed and 
fitted SEDs, respectively, and the filled symbols and dashed lines 
are those for the extinction-corrected values adopting 
the interstellar reddening law of $\rv = 3.1$ \citep{draine03}.
The dotted line of Object 1 is a model SED
assuming a binary system comprising two early-type stars 
of $T=$24,000~K (see \S~4). 
Object 2, which does not have enough data points, 
is excluded from the fitting. \label{fig8}}
\end{figure}

A comparison of the observed and best-fit SEDs of all the IR-excess
stellar objects other than Object 2 is presented in
Figure~\ref{fig8}. We note that the observed {\it R}-band fluxes of
Objects 5, 10, and 11 are somewhat lower than expected from the
model calculations. The origin of these discrepancies in the
$R$-band is not clear, although the slightly loose photometric
calibration in the USNO-B1.0 catalog and/or the conversion between
the USNO photographic magnitude system \citep{usno} and the standard
Johnson-Cousin system may be responsible. 
Table~\ref{tbl4} contains the resulting best-fit parameters 
derived in the SED fits with their reduced chi-square values.
The spectral types are taken from models of \citet{martins05} (for
O-type stars) and \citet{schmidt} (for B-type stars), which are
mapped to the best-fit temperatures of the Kurucz models. All the
IR-excess stellar objects are early-type stars of similar spectral
type between O9 and B2.5, which is consistent with the previous
results from their {\it JHK}$_s$ colors and also with the results
from the NIR spectral analyses.

The best-fit SED temperatures of the IR-excess stellar objects in
Table~\ref{tbl4} are similar to those derived from the {\it EW}s of
their NIR absorption lines (\S~3.2), except for those of Objects 1
and 5 which show relatively large discrepancies: 26,000~K (from the
{\it EW}s) in contrast to 32,000~K (from the SED fits) for Object 1,
and 33,000~K (from the {\it EW}s) in contrast to 27,000~K (from the
SED fits) for Object 5. As we explain in \S~3.2, we consider this
temperature difference of $\sim$6,000~K, which is roughly equivalent
to two spectral subclasses, to lie within the uncertainty range
inherent in the classification of the spectral types of the
IR-excess stellar objects. Comparing Objects 1 and 5, however, we
note that the former has stronger H absorption lines, which would
indicate a lower temperature (Figures~\ref{fig3} and \ref{fig4}),
even though it has a higher SED temperature (Figure~\ref{fig8} and
Table~\ref{tbl4}). One possible explanation for this is that Object
1 is a binary system of two early-type stars, similar to Object 2.
The dotted line of Object 1 in Figure~\ref{fig8} shows such an
example where we calculate the expected SED of a binary system
comprising two early-type stars of $T$ = 24,000~K.

Table~\ref{tbl4} also contains the extinctions of the IR-excess
stellar objects, where the six objects (1, 5, 7, 8, 10, and 11) with
the obtained NIR spectra show almost identical extinctions to those
in Table~\ref{tbl3}. The extinctions of all the IR-excess stellar
objects are between $\av$ = 7 and $\av$ = 11 mag, and Objects 3, 4,
6, 7, and 9, which have extinctions of $\av \gtrsim 8.0$ mag, appear
to be located in the western part of the IR loop
(Figure~\ref{fig1}). This is consistent with their relative
faintness as listed in Table~\ref{tbl1} and their non-detection in
the {\it B}-band, which is the most sensitive to increased
extinctions. The increased extinctions indicate a potential
enhancement of matter in the western part of the IR loop, and the
relatively smaller ($\av \sim$7 mag) extinctions of the objects in
the outer part of the loop, i.e., Objects 8, 10, and 11, support
this interpretation.

\citet{lu02} previously estimated the average X-ray absorption
column density toward the central pulsar in G54.1+0.3 to be 
$N({\rm H}) = (1.6 \pm 0.1) \times 10^{22}$ cm$^{-2}$ (or $\av=8.4 \pm 0.5$ mag), 
whereas in the deeper X-ray observations of \citet{temim10}, 
a spatial variation in the absorption column density with a mean value of 
$N({\rm H}) = (1.95 \pm 0.04) \times 10^{22}$ cm$^{-2}$ (or $\av=10.2 \pm 0.2$ mag) 
was found. The extinction values estimated from the X-ray observations are 
somewhat larger than those of the IR-excess stellar objects 
from the SED fits in this study, which
suggests the possibility of a much wider dust distribution than the
observed IR loop, e.g., a shell-like distribution surrounding the
X-ray PWN of G54.1+0.3 that is invisible due to the absence of
heating sources.

\section{Discussion}

Two possible scenarios that explain the relation between the
IR-excess stellar objects and the SNR G54.1+0.3 have been proposed:
(1) \citet{koo08} proposed that the IR-excess stellar objects are 
massive young stellar objects (YSOs) 
whose formation was triggered by the progenitor of G54.1+0.3 
based on their SEDs and spatial confinement to the IR loop; and 
(2) \citet{temim10}, using the results obtained with {\it Spitzer} IRS
observations, proposed that the IR-excess stellar objects are
members of a stellar cluster to which the progenitor of the SNR
originally belonged and that the IR emission is from the ejecta dust
heated by the stellar objects. We discuss these two scenarios in the
context of the results presented in this paper.

\subsection{Young Stellar Objects Triggered by the Progenitor Star}

In the first scenario, the IR-excess stellar objects are interpreted
to be massive YSOs whose formation was triggered by the progenitor
of the SNR G54.1+0.3 \citep{koo08}. As \citet{koo08} has previously
shown, the SEDs of the IR-excess stellar objects have a dip at
around 8~$\micron$ with a strong MIR excess, similar to those of
stars without substantial circumstellar material such as YSOs with a
truncated disk or Herbig Ae/Be stars with a breakup inner disk
\citep{hillenbrand92,malfait98,dalessio05}.
However, the NIR spectra (Figures~\ref{fig3} and \ref{fig4}) 
of the IR-excess stellar objects
do not show any emission lines that are often found in Herbig Ae/Be
stars \citep{hillenbrand92,malfait98}. This indicates that the
IR-excess stellar objects are in a relatively later evolutionary
stage of pre-MS stars with only little nearby circumstellar material
remaining. In general, the disk lifetime of YSOs is well correlated
with the mass of the central star, and for stars more massive than
7\,$M_{\sun}$, it is known that disk dispersion takes less than 1 Myr
\citep{hillenbrand92,alonso09}. Hence, the IR-excess stellar objects
are likely to be a few million years old, and their formation may
have been triggered during the late MS or post-MS evolutionary
stages of the progenitor. In the case of the later, a potential
delay in the triggered formation process due to local
environment effects has been proposed \citep{koo08}.

The IR loop in G54.1+0.3, where the IR-excess stellar objects are
embedded, was suggested to be composed of fast-expanding SN ejecta
dust \citep{temim10}, although this does not preclude the existence
of the local ISM and/or circumstellar material in the loop. If the
ejecta dust is illuminated by the IR-excess stellar objects, then
significant enhancement of the MIR emission can occur, and
depending on the distance between the two, the 8~$\micron$ dip
features seen in the SEDs of the latter (Figure~\ref{fig8}) may
arise. If the IR emission of the stellar objects is from SN dust
rather than circumstellar dust, 
then the IR-excess stellar objects would be older than a few million years, 
and subsequently, the triggered formation by the progenitor star 
of G54.1+0.3 could have happened earlier, e.g., during their MS stages.

On the other hand, we note that confirmation of the IR-excess stellar objects 
as a result of triggered star formation is somewhat uncertain based on the available
information from the existing observations. As described above, the spatial 
confinement of the IR-excess stellar objects in the IR loop surrounding the 
PWN in G54.1+0.3 indicates potential triggered star formation by the SN 
progenitor star. However, recent numerical simulations of the feedback effects
on star formation \citep{dale11,dale12} suggest that the features such as 
bubble walls or pillars, in which young stars are embedded, do not always trace 
triggered star formation, although they may serve as a good indicator. 
According to the results, such geometrical distributions of stars can be 
produced without feedback, depending on environments \citep{dale11,dale12}.

\subsection{Stellar Cluster Origin}

According to the second scenario, an SN exploded as a member of a
cluster of stars that includes the IR-excess stellar objects, and
the stellar cluster illuminates the SN ejecta dust to produce the
strong MIR emission that is observed \citep{temim10}.
In this case,
the results of the NIR spectroscopy reported in this paper help
constrain the mass and type of the SN progenitor as follows. First,
the mass ($M \sim 17\,M_{\sun}$) of an O9 star, which is the
earliest spectral type of the IR-excess stellar objects
(Table~\ref{tbl4}), gives the lower limit of the progenitor mass,
and considering the continuous distribution of the initial mass
function \citep[e.g.,][]{salpeter55}, the progenitor mass cannot be
much greater than that of an O9 star. In other words, it is unlikely
that the progenitor was massive enough to be a Wolf-Rayet star of
$\gtrsim$35\,$M_{\sun}$ \citep[e.g.,][and the references
therein]{smith11}. Consequently, we can constrain the progenitor
mass to be in the range of 18--35\,$M_{\sun}$, although the upper
limit may be significantly lower than 35\,$M_{\sun}$.

The SN type of G54.1+0.3 has not yet been definitively confirmed.
\citet{chevalier05} previously ruled out the IIL/b possibility due
to a lack of any observational evidence indicating the existence of
strong circumstellar interactions and suggested that G54.1+0.3 is an
SN IIP or Ib/c. If the progenitor star was a single star, then we
believe that it is less likely to be an SN Ib/c, given the upper
mass limit, which leaves us only with the IIP possibility. The mass
range of SN IIP progenitors was first calculated to be
9--25\,$M_{\sun}$ by \citet{heger03} based only on single star
evolutionary models. \citet{smith11}, however, recently suggested a
smaller range of 8--18\,$M_{\sun}$ after observationally accounting
for contributions from both single stars and binaries in the entire
core-collapse populations. This range of \citet{smith11} for the
progenitor mass of SNe IIP places the G54.1+0.3 progenitor mass
close to the upper limit, similar to that of the O9 star, which is
the earliest existing member of the cluster as noted. Therefore, if
the progenitor of G54.1+0.3 was indeed a member of the cluster to
which the IR-excess stellar objects also belong and exploded as an SN
IIP, the spectral type of the progenitor should be that of a late O-type star
(e.g., O8). Finally, we note that in principle we cannot yet
completely rule out the possibility of the G54.1+0.3 progenitor
being a binary that exploded as an SN Ib/c, although we believe that
this possibility is low since SNe Ib/c from binaries are much less
abundant than SNe IIP.

\section{Summary and Conclusions}

We present the results of broadband NIR spectroscopic
observations and SED analyses of the IR-excess stellar objects
discovered in the IR loop of the SNR G54.1+0.3. A summary of our
results and main conclusions are below.

(1) We spectroscopically determine the spectral types of the
six IR-excess stellar objects to be between O9 and B2 based on the
empirical relation (see Eqn.~\ref{eqn1}) between the {\it EW}s of
the H lines and stellar temperatures that we establish using the
results of previous studies available in the literature. The
determined spectral types are consistent with those from 
the {\it JHK}$_s$ color analyses. The established relation between 
the {\it EW}s of the H lines and stellar temperatures can generally be used
to determine the spectral types of OB stars with NIR spectra, and it
is worthwhile to refine the relation in the future by increasing the
sample size.

(2) We determine the distance, which is independent of the Galactic
rotation model, to the IR-excess stellar objects to be 6.0 $\pm$ 0.4
kpc based on our SED fits. The distance of 6 kpc is somewhat smaller
than but still compatible to the previously proposed distances to
G54.1+0.3.

(3) The photometric spectral types of most of the IR-excess stellar
objects lie within the range of O9--B2.5, which is consistent with
the spectroscopic results. The extinctions of the IR-excess stellar
objects, with a mean of $\av$ =  7.9 $\pm$ 0.1 mag, increase in the
western part of the IR loop, indicating an enhancement of matter in
that region.

(4) If the formation of the IR-excess stellar objects was triggered
by the progenitor star of G54.1+0.3, then the lack of emission lines
in our NIR spectra indicates that the objects are likely to be a few
million years old, which is suggestive of triggering occurring
during the late MS or post-MS stages of the progenitor. However, the
potential contributions of the SN ejecta dust to the observed MIR
emission of the IR-excess stellar objects may also be consistent
with triggering during the MS stages.

(5) If the IR-excess stellar objects are members of a stellar
cluster to which the progenitor originally belonged, then our NIR
spectroscopic results constrain the mass of the progenitor to be
slightly greater than 17\,$M_{\sun}$. In such a case, it is more
likely that G54.1+0.3 was an SN IIP.

The NIR spectroscopic observations of the IR-excess stellar objects
in the SNR G54.1+0.3 presented in this study provide important
information about their nature, particularly the spectral types.
However, their origin and relation with the SNR are still
unclear. Further investigations on the dust in the IR loop and the
progenitor of G54.1+0.3, accompanied by the use of physical models,
will help decipher the nature of the IR-excess stellar objects and
substantiate a more plausible scenario of their origin.

This publication makes use of data products from 
the Two Micron All Sky Survey, which is a joint project of 
the University of Massachusetts and the Infrared Processing and 
Analysis Center/California Institute of Technology, funded by 
the National Aeronautics and Space Administration and 
the National Science Foundation. 
This work was supported by NRF(National Research Foundation of Korea) 
Grant funded by the Korean Government
(NRF-2012-Fostering Core Leaders of the Future Basic Science Program).
B.-C. K. is supported by Basic Science Research program 
through the National Research Foundation of Korea (NRF) funded by 
the Ministry of Education, Science and Technology (NRF-2011-0007223).
D.-S. M. acknowledges support from the Natural Science and 
Engineering Research Council of Canada. This paper was
studied with the support of the Ministry of Education
Science and Technology (MEST) and the Korean 
Federation of Science and Technology Societies (KOFST).
We thank Prof. Sang-Gak Lee for helpful advice on 
calibration and analysis of TripleSpec spectra.

\clearpage

\clearpage

\clearpage

\clearpage

\clearpage

\begin{deluxetable}{cccccccccccccccccc}
\rotate
\tabletypesize{\tiny}
\tablecaption{
Coordinates and magnitudes of the IR-excess stellar objects 
in the SNR G54.1+0.3 \label{tbl1}}
\tablewidth{0pt}
\tablehead{
\colhead{Object} & \colhead{Ra (J2000)} & \colhead{Dec (J2000)} & 
\colhead{\it B} & \colhead{\it V} & \colhead{\it R} & \colhead{\it I} &
\colhead{\it J} & \colhead{\it H} & \colhead{\it K$_s$} &
\colhead{3.6} & \colhead{4.5} & \colhead{5.8} & \colhead{8.0} &
\colhead{MIPS 24} & \colhead{MIPS 70} & \colhead{{\em AKARI} L15} & 
\colhead{{\em AKARI} L24}
}
\startdata
1 & 19:30:30.38 & 18:51:30.64 & 
19.61 & 17.68 & 16.70 & 15.82 &
12.81 & 12.05 & 11.74 & 
11.35 & 11.24 & 10.98 & 9.01 & 
2.34 & 0.34 & 5.16 & 2.31 \\
&&& 0.50 & 0.50 & 0.50 & 0.50 & 0.02 & 0.03 & 0.02 &
0.03 & 0.04 & 0.05 & 0.03 & 0.06 & 0.33 & 0.05 & 0.06\\
2 & 19:30:31.29 &18:51:32.81 &
... & ... & ... & ... & 
... & ... & ... &
12.41 & 12.04 & 11.39 & 9.70 &
3.40 & 0.45 & 5.85 & 3.12 \\
&&& ... & ... & ... & ... & ... & ... & ... &
0.16 & 0.12 & 0.12 & 0.03 & 0.08 & 0.53 & 0.06 & 0.07\\
3 & 19:30:27.88 & 18:52:35.28 &
... & ... & ... & ... & 
14.53 & 13.64 & 13.20 &
12.80 & 12.62 & ... & 10.82 &
3.40 & 0.01 & 6.44 & 3.34\\
&&& ... & ... & ... & ... & 0.04 & 0.03 & 0.04 &
0.05 & 0.07 & ... & 0.05 & 0.08 & 0.41 & 0.06 & 0.07\\
4 & 19:30:28.61 & 18:51:52.37 &
... & ... & 17.86 & 17.06 &
13.93 & 13.09 & 12.71 &
12.30 & 12.24 & 12.14 & 11.64 & 
3.43 & 0.46 & 6.94 & 3.51\\
&&& ... & ... & 0.50 & 0.50 & 0.03 & 0.03 & 0.03 &
0.04 & 0.06 & 0.12 & 0.07 & 0.08 & 0.35 & 0.06 & 0.08\\
5 & 19:30:29.29 & 18:51:37.43 &
20.43 & ... & 17.61 & 16.48 & 
13.32 & 12.54 & 12.17 &
11.95 & 11.86 & 11.74 & 9.87 & 
3.66 & 0.60 & 7.02 & 4.29\\
&&& 0.50 & ... & 0.50 & 0.50 & 0.02 & 0.02 & 0.02 &
0.04 & 0.05 & 0.08 & 0.03 & 0.07 & 0.75 & 0.07 & 0.08\\
6 & 19:30:27.58 & 18:51:50.31 &
... & ... & ... & ... &
13.61 & 12.75 & 12.11 &
11.47 & 11.08 & 10.73 & 9.60 &
3.74 & 0.25 & 6.40 & 4.16\\
&&& ... & ... & ... & ... & 0.05 & 0.04 & 0.04 &
0.03 & 0.03 & 0.05 & 0.03 & 0.07 & 0.47 & 0.04 & 0.09\\
7 & 19:30:26.69 & 18:52:36.89 &
... & ... & 18.11 & 17.04 &
13.70 & 12.83 & 12.42 &
12.07 & 11.96 & 11.92 & 11.13 &
3.79 & ... & 6.73 & 4.24 \\
&&& ... & ... & 0.50 & 0.50 & 0.03 & 0.03 & 0.03 &
0.04 & 0.05 & 0.07 & 0.05 & 0.09 & ... & 0.06 & 0.14\\
8 & 19:30:25.29 & 18:51:53.69 & 
19.92 & ... & 16.66 & 15.76 &
13.40 & 12.61 & 12.25 &
12.01 & 11.80 & 11.86 & 11.06 &
3.99 & 0.22 & 7.24 & 4.16 \\
&&& 0.50 & ... & 0.50 & 0.50 & 0.03 & 0.03 & 0.03 &
0.04 & 0.06 & 0.08 & 0.05 & 0.04 & 0.93 & 0.07 & 0.16\\
9 & 19:30:29.31 & 18:52:34.77 & 
... & ... & ... & ... &
14.31 & 13.40 & 12.93 &
12.60 & 12.46 & 12.29 & ... &
4.74 & 2.46 & 7.84 & 4.79 \\
&&& ... & ... & ... & ... & 0.03 & 0.03 & 0.03 &
0.05 & 0.06 & 0.10 & ... & 0.09 & 4.35 & 0.10 & 0.09\\
10 & 19:30:29.72 & 18:53:18.27 &
20.50 & ... & 18.14 & 16.87 &
13.65 & 12.84 & 12.50 &
12.22 & 12.15 & 11.92 & 11.98 &
5.33 & 2.39 & 8.86 & 5.06 \\
&&& 0.50 & ... & 0.50 & 0.50 & 0.03 & 0.03 & 0.03 &
0.03 & 0.06 & 0.07 & 0.08 & 0.10 & 1.82 & 0.07 & 0.05\\
11 & 19:30:31.42 & 18:52:48.52 &
20.47 & ... & 17.84 & 16.91 & 
13.83 & 13.04 & 12.66 &
12.39 & 12.36 & 12.21 & 11.59 &
5.75 & 3.58  & 8.52  & 5.27 \\
&&& 0.50 & ... & 0.50 & 0.50 & 0.03 & 0.04 & 0.04 &
0.03 & 0.06 & 0.09 & 0.06 & 0.10 & 2.92 & 0.06 & 0.04
\enddata
\tablecomments{The second rows list the magnitude errors (mag).}
\tablecomments{The {\it BRI} magnitudes (Cols. 4, 6, \& 7) are from 
the USNO-B1.0 catalog \cite{usno}, and the {\it V} magnitude (Col. 5) is from 
the NOMAD catalog \cite{nomad}. 
The {\it JHK}$_s$ magnitudes (Cols. 8 to 10) and 
the IRAC magnitudes (Cols. 11 to 14) are from the 2MASS All-Sky PSC 
and the GLIMPSE catalog, respectively.
The {\it Spitzer} MIPS and {\em AKARI} MIR magnitudes are determined 
by PSF photometry \citep{koo08}.} 
\end{deluxetable}

\clearpage

\begin{deluxetable}{ccccccccccc}
\rotate
\tabletypesize{\tiny}
\tablecaption{
Equivalent widths of the identified lines \label{tbl2}}
\tablewidth{0pt}
\tablehead{
\colhead{ } & \colhead{\heI} & \colhead{Pa$\beta$} & \colhead{\heI} & 
\colhead{Br14} & \colhead{Br13} & \colhead{Br12} & \colhead{Br11} & 
\colhead{\heI} & \colhead{\heI} & \colhead{Br$\gamma$}\\
\colhead{Object} & \colhead{1.279$\micron$} & \colhead{1.282$\micron$} & 
\colhead{1.297$\micron$} & \colhead{1.589$\micron$} & 
\colhead{1.611$\micron$} & \colhead{1.641$\micron$} & 
\colhead{1.681$\micron$} & \colhead{1.701$\micron$} & 
\colhead{2.113$\micron$} & \colhead{2.166$\micron$}
}
\startdata
1 & 2.13 $\pm$ 0.29 & 3.53 $\pm$ 0.30 & 0.77 $\pm$ 0.05 & 
4.87 $\pm$ 0.82 & 5.36 $\pm$ 0.75 & 3.69 $\pm$ 0.55 & 3.39 $\pm$ 0.56 & 
0.81 $\pm$ 0.25 & 1.19 $\pm$ 0.69 & 6.61 $\pm$ 1.21\\
5 & 2.10 $\pm$ 0.37 & 3.14 $\pm$ 0.38 & 0.56 $\pm$ 0.25 & 
4.84 $\pm$ 1.57 & 3.24 $\pm$ 0.94 & 2.55 $\pm$ 0.73 & 1.77 $\pm$ 0.65 & 
1.00 $\pm$ 0.30 & 0.78 $\pm$ 0.45 & 7.41 $\pm$ 1.80\tablenotemark{a}\\
7 & 1.81 $\pm$ 0.53 & 3.91 $\pm$ 0.54 & 0.31 $\pm$ 0.28 & 
6.39 $\pm$ 1.02 & 6.88 $\pm$ 0.82 & 3.70 $\pm$ 0.57 & 5.28 $\pm$ 0.69\tablenotemark{a} & 
0.74 $\pm$ 0.30 & 0.51 $\pm$ 0.60 & 4.97 $\pm$ 1.13\\
8 & 1.54 $\pm$ 0.31 & 3.60 $\pm$ 0.34 & 0.60 $\pm$ 0.27 & 
4.17 $\pm$ 0.60 & 5.82 $\pm$ 0.62 & 3.07 $\pm$ 0.41 & 3.97 $\pm$ 0.52 & 
0.75 $\pm$ 0.25 & 1.80 $\pm$ 0.89 & 3.40 $\pm$ 0.89\\
10 & 1.47 $\pm$ 0.40 & 3.46 $\pm$ 0.46 & 0.35 $\pm$ 0.22 & 
5.68 $\pm$ 1.06 & 6.30 $\pm$ 0.94 & 3.49 $\pm$ 0.63 & 4.14 $\pm$ 0.77 & 
1.22 $\pm$ 0.30 & 1.29 $\pm$ 1.02 & 4.54 $\pm$ 1.44\\
11 & 1.77 $\pm$ 0.46 & 3.59 $\pm$ 0.56 & 1.29 $\pm$ 0.51\tablenotemark{a} & 
5.83 $\pm$ 1.16 & 7.73 $\pm$ 1.14 & 2.97 $\pm$ 0.73 & 
3.99 $\pm$ 0.85 & 0.79 $\pm$ 0.43 & 0.67 $\pm$ 0.80 & 6.26 $\pm$ 1.92\\
\enddata
\tablenotetext{a}{Since these lines or the continuum of these lines are 
contaminated by telluric lines or noises, the equivalent widths of 
these lines might be measured larger than expected, 
so these values are upper limits.}
\tablecomments{The central wavelengths are the vacuum wavelengths, 
and the {\it EW}s are in \AA.}
\end{deluxetable}

\clearpage

\begin{table}
\begin{center}
\caption{
Stellar parameters derived from the relation between the equivalent widths
of Br11 and the stellar temperatures \label{tbl3}}
\begin{tabular}{ccccc}
\tableline\tableline
Object & Temperature\tablenotemark{a} & Spectral & 
Distance & $\av$ \\
     & (K)  & Type & (kpc) & (mag) \\
\tableline
1 & 26000 $\pm$ 2000 & B1 (B0.5-B1.5) & 
$4.6^{+0.4}_{-0.3}$ & $7.4^{+\lesssim 0.05}_{-0.3}$ \\
5 & 33000 $\pm$ 3000 & O9 (O7.5-B0) &
$8.1^{+1.9}_{-1.3}$ & $7.5^{+\lesssim 0.05}_{-0.1}$ \\ 
7 & $\gtrsim$ 17000 & $\lesssim$ B3.5 &
$\gtrsim$ 3.3 & $\gtrsim$ 7.8 \\
8 & 23000 $\pm$ 2000 & B1.5 (B1-B2) &
$5.4^{+0.5}_{-0.7}$ & $6.9^{+0.1}_{-\lesssim 0.05}$ \\ 
10 & 22000 $\pm$ 3000 &  B2 (B1-B3) &
$5.5^{+0.9}_{-1.2}$ & $7.4^{+0.1}_{-0.2}$ \\ 
11 & 23000 $\pm$ 4000 & B1.5 (B1-B2.5) &
$6.5^{+1.2}_{-1.7}$ & $7.1^{+0.2}_{-0.2}$ \\ 
\tableline
\end{tabular}
\tablenotetext{a}{The effective temperatures of the MS stars are 
adopted from \cite{martins05} for O-type stars and those of \cite{schmidt} 
for stars later than B0.}
\end{center}
\end{table}

\begin{table}
\begin{center}
\caption{
Spectral types and extinctions derived in the SED fits using 
photometric observations with fixed distance of 6 kpc 
\label{tbl4}}
\begin{tabular}{ccccc}
\tableline\tableline
Object & Temperature\tablenotemark{a,b} & Spectral &
$\av$  & $\chi^2_{red}$ \\
      &  (K)  & Type  & (mag) &  \\
\tableline
   1  & 32000  &  O9.5  &    7.4 $\pm$ 0.1  & 2.7  \\
   2  & ...    &  ...   &    ...          & ...   \\
   3  & 20000  &  B2.5  &    8.0 $\pm$ 0.2  & 2.3  \\
   4  & 23000  &  B1.5  &    7.6 $\pm$ 0.2  & 3.0  \\
   5  & 27000  &  B0.5  &    7.4 $\pm$ 0.1  & 13.0  \\
   6  & 33000  &    O9  &   11.0 $\pm$ 0.2  & 11.0  \\
   7  & 25000  &    B1  &    8.0 $\pm$ 0.1  & 6.8  \\
   8  & 26000  &    B1  &    7.0 $\pm$ 0.2  & 4.0  \\
   9  & 21000  &    B2  &    8.1 $\pm$ 0.2  & 6.6  \\
  10  & 24000  &  B1.5  &    7.3 $\pm$ 0.1  & 12.2  \\
  11  & 22000  &    B2  &    7.1 $\pm$ 0.1  & 6.0  \\
\tableline
\end{tabular}
\tablenotetext{a}{The effective temperatures of the MS stars are 
adopted from \cite{martins05} for O-type stars and those of \cite{schmidt} 
for stars later than B0.}
\tablenotetext{b}{The 1-sigma uncertainty of temperature is 
less than temperature interval in the fitting.}
\end{center}
\end{table}

\end{document}